\documentclass[aps,prb,twocolumn,groupedaddress,floatfix,showpacs]{revtex4}

\usepackage{graphicx}% Include figure files
\usepackage{dcolumn} % Align table columns on decimal point
\usepackage{bm}      % bold math
\usepackage{amsmath} % \text{}

\begin{document}

\title{Charge order and spin-singlet pairs formation in Ti$_4$O$_7$}

\author{I.~Leonov$^{1}$\email[e-mail: ]{Ivan.Leonov@Physik.uni-Augsburg.de},
A.~N.~Yaresko$^2$, V.~N.~Antonov$^3$, U.~Schwingenschl\"ogl$^4$, V.~Eyert$^4$, and V.~I.~Anisimov$^5$}

\affiliation{$^1$ Theoretical Physics III, Center for Electronic Correlations 
and Magnetism, Institute for Physics, University of Augsburg, Germany}
\affiliation{$^2$ Max-Planck Institute for the Physics of Complex Systems, 
Dresden, Germany}
\affiliation{$^3$ Institute of Metal Physics, Vernadskii Street, 03142 Kiev, 
Ukraine}
\affiliation{$^4$ Theoretical Physics II, Institute for Physics, 
University of Augsburg, Germany}
\affiliation{$^5$ Institute of Metal Physics, Russian Academy of Science-Ural 
Division, 620219 Yekaterinburg GSP-170, Russia}

\date{\today}

\begin{abstract}
Charge ordering in the low-temperature triclinic structure of titanium 
oxide (Ti$_4$O$_7$) is investigated using the local density approximation 
(LDA)+$U$ method. Although the total $3d$ charge separation is rather 
small, an orbital order parameter defined as the difference between 
$t_{2g}$ occupancies of Ti$^{3+}$ and Ti$^{4+}$ cations is large and 
gives direct evidence for charge ordering. Ti $4s$ and $4p$ states make 
a large contribution to the static ``screening'' of the total $3d$ charge 
difference. This effective charge screening leads to complete loss of the 
disproportionation between the charges at 3+ and 4+ Ti sites. The occupied 
$t_{2g}$ states of Ti$^{3+}$ cations are predominantly of $d_{xy}$ character 
and form a spin-singlet molecular orbital via strong direct 
antiferromagnetic exchange coupling between neighboring Ti(1) 
and Ti(3) sites, whereas the role of superexchange is found 
to be negligible.

\end{abstract}


\maketitle

%%%%%%%%%%%%%%%%%%%%%%%%%%%%%%%%%%%%%%%%%%%%%%%%%%%%%%%%%%%%%%%%%%%%%%%%%%%%%%%%%
% Introduction
%%%%%%%%%%%%%%%%%%%%%%%%%%%%%%%%%%%%%%%%%%%%%%%%%%%%%%%%%%%%%%%%%%%%%%%%%%%%%%%%%

\section{Introduction}
\label{sec:introd}

The mixed valent transition metal oxides, that simultaneously 
contain metal atoms in two (or more) different valence states, are of 
strong current interest.\cite{Coey04} One of the classical examples of 
such a system is magnetite (Fe$_3$O$_4$), in which a 
first-order metal-insulator transition occurs at $\sim$120\,K.\cite{Rev01} 
According to Verwey, this transition is caused by the spatial ordering of 
2+ and 3+ Fe cations on the octahedral $B$-sublattice of the inverted 
spinel structure AB$_2$O$_4$.\cite{V39,VHR47} Recently, a local spin 
density approximation (LSDA)+$U$ study of the low-temperature phase of 
Fe$_3$O$_4$ resulted in a charge and orbitally ordered insulating ground 
state with a well pronounced orbital order.\cite{LYA04,JGH04} However, 
the strong difference in $t_{2g}$ occupancies of 2+ and 3+ Fe was found 
to be drastically reduced by effective ``static'' screening.\cite{screening} 
A similar result (see Ref.~\onlinecite{LYA05}) has been obtained for 
another iron oxide, containing both 2+ and 3+ Fe cations, iron oxoborate 
(Fe$_2$OBO$_3$), which shows a broad semiconductor-to-semiconductor 
transition at $\sim$317\,K associated (as in Fe$_3$O$_4$) with a 
spatial order-disorder transformation of 2+ and 3+ Fe cations on 
quasi-one-dimensional Fe chains.\cite{ACP92,ABRM98,ABRM99} 

The aforementioned phenomena of sharp metal-insulator transitions
associated with pronounced charge and/or orbital ordering are
characteristic for the \textit{Magn\'eli} phases $ {\rm M_nO_{2n-1}} $ 
($ {\rm M = Ti, V} $).\cite{V7O13} These compounds form a homologous 
series and have been studied recently to understand the differences 
in crystal structures and electronic properties between the end 
members $ {\rm MO_2} $ ($ {\rm n \to \infty} $) and $ {\rm M_2O_3} $ 
($ {\rm n = 2} $).\cite{schwingen03} In particular, the metal-insulator 
transition of $ {\rm VO_2} $ discovered some fifty years ago still is 
the subject of ongoing controversy and is another ``hot topic'' in 
solid state physics. LDA calculations have revealed strong influence 
of the structural degrees of freedom on the electronic properties of 
$ {\rm VO_2} $ and neighbouring rutile-type 
dioxides.\cite{wentz94a,KSA02,eyert02b,eyert00,eyert02a} 
In this scenario the characteristic dimerization and antiferroelectric 
displacement of the metal atoms translate into orbital ordering within 
the $ t_{2g} $ states and a Peierls-like singlet formation between 
neighbouring sites. Recently, this was confirmed by LDA+DMFT calculations, 
which suggested to regard the transition of $ {\rm VO_2} $ as a 
correlation-assisted Peierls-transition.\cite{biermann05} 

Ti$_4$O$_7$ titanium oxide is another remarkable member of the
\textit{Magn\'eli} phases with $n$=4 which shows metal-insulator transitions
associated with the spatial charge ordering.
It is a mixed valent compound which has an even mixture of 3+ and 4+ Ti
cations (Ti$_2^{3+}$Ti$_2^{4+}$O$_7$), corresponding to an average $3d$
occupation of 1/2 electron per Ti site.
Electrical resistivity, specific heat, magnetic susceptibility, and x-ray
diffraction data reveal two first-order transitions in the temperature
range of 130-150\,K.\cite{BF69,LSC76,MMD72} At 150\,K a metal-semiconductor
transition occurs without measureable hysteresis in resistivity and
specific heat. It is followed by a semiconductor-semiconductor transition
at 130-140\,K, which again is characterized by an almost two orders of
magnitude abrupt increase in electrical resistivity and has a hysteresis of
several degrees.  \cite{BF69,LSC76} The magnetic susceptibility shows a
sharp enhancement when heating through 150\,K. However, it is small and
temperature independent below this temperature and does
not show any anomaly at 140\,K.

\begin{figure}[tbp!]
\centerline{\includegraphics[height=0.5\textheight,clip]{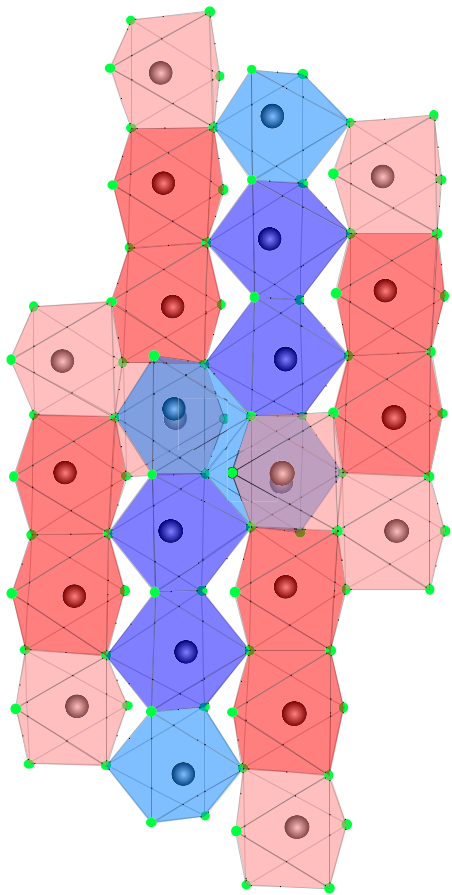}}
%\centerline{\includegraphics[width=0.24\textwidth,clip,angle=90]{Ti4O7_struc.eps}}
\caption{\label{fig:struc}
(color online) The low-temperature crystal structure of Ti$_4$O$_7$.
Chains of four Ti sites run parallel to the pseudo-rutile $c$-axis.
Red and blue (light and dark on the black and white image) chains of four Ti 
atoms correspond to the (a) and (b) chains Ti atoms, respectively. Further 
gradation of red and blue on light and dark subsets indicates inequivalent
Ti sites in (a) and (b) chains.
}
\end{figure}

The crystal structure of Ti$_4$O$_7$ (see Fig.~\ref{fig:struc}) can be
viewed as rutile-type slabs of 
infinite extension and four Ti sites thickness, separated by shear planes
with a corundum-like atomic arrangement.  Below 130\,K it crystallizes in a
triclinic crystal structure with two formula units per primitive unit
cell.\cite{MD71,HM79,LM84} Four crystallographically inequivalent Ti
sites are found at the centers of distorted oxygen octahedra. They
form two types of chains, namely, (a) 1-3-3-1 and (b) 2-4-4-2, which run
parallel to the pseudo-rutile $c$-axis and are separated by the
crystallographic shear planes. Although interatomic distances in the
(b)-chain are almost uniform (3.01 and 3.07 \AA\ between 4-4 and 2-4 Ti 
sites, respectively) they are remarkably different for the (a)-chain 
(3.11 and 2.79 \AA\ between 3-3 and 1-3 Ti sites).

%%%%%%%%%%%%%%%%%%%%%%%%%%%%%%%%%%%%%%%%%%%%%%%%%%%%%%%%%%%%%%%%%%%%%%%%%%%%%%%%%
%%%%%%%%%%%%%%%%%%%%%%%%%%%%%%%%%%%%%%%%%%%%%%%%%%%%%%%%%%%%%%%%%%%%%%%%%%%%%%%%%

Accurate determination of the crystal structure allowed to elucidate 
the nature of the three phases distinguished by the two first-order 
transitions.\cite{LM84,MD71,HM79,MMD72} In particular, in the 
metallic phase the average Ti-O bond lengths for crystallographically 
inequivalent TiO$_6$ octahedra are very similar which results in the 
average valence state of 3.5+ per each Ti cation. Below 130\,K charge 
has been transferred from the (b) to the (a)-chains.
In addition, Ti$^{3+}$ cations in alternate (a)-chains are paired to 
form nonmagnetic metal-metal bonds, whereas in the intermediate 
phase pairing also persists but its long-range order calls for 
a fivefold supercell.\cite{LSC76} Thus, the 130-140\,K transition 
is associated with a transition to the phase with a long-range 
order of Ti$^{3+}$-Ti$^{3+}$ pairs, whereas above 150 K 3+ 
and 4+ Ti cations are disordered. The presence of the 
Ti$^{3+}$-Ti$^{3+}$ pairs strongly differentiates Ti$_4$O$_7$ from 
Fe$_3$O$_4$ and results in two steep first-order transitions found 
in the electrical resistivity.

Recent LDA band structure calculations of both high- and low-temperature
phase of Ti$_4$O$_7$ results in significant $t_{2g}$ charge separation 
between crystallographically independent 3+ and 4+ Ti sites in the 
low-temperature phase, whereas a rather isotropic occupation of the 
$t_{2g}$ states has been found at room-temperature.\cite{ESE04} 
While, in addition, an orbital order at the Ti $d^1$ chains originating 
from metal-metal dimerization was found, the LDA gave 
only metallic solution with semimetallic-like band overlap instead 
of the semiconducting gap. This problem is overcome in our work 
taking into account strong electronic correlations in Ti $3d$
shell using the LDA+$U$ method.

In the present paper we investigate the electronic structure of the 
low-temperature phase using the LDA+$U$ approach in the 
tight-binding linear muffin-tin orbital (TB-LMTO) calculation 
scheme.\cite{ole,AZA91,LAZ95} 
The LDA+$U$ calculations result in a charge and orbitally 
ordered insulator with an energy gap of 0.29\,eV, which is in 
a good agreement with an experimental gap value of 0.25\,eV.
From our results, we propose an 
orbital order parameter, defined as the difference between 
$t_{2g}$ majority/minority spin occupancies of 
Ti(1)$^{3+}$/Ti(3)$^{3+}$ and Ti(2)$^{4+}$/Ti(4)$^{4+}$ cations, 
respectively. This order parameter is found to be quite large, although 
the total $3d$ charge difference between 3+ and 4+ cations, remains small. 
Also it is interesting to note that the total charge separation between 
3+ and 4+ Ti cations completely is lost due to 
efficient screening by the rearrangement of the other Ti electrons.
In addition, we find a strong antiferromagnetic coupling of 
$ J \approx 1700 $\,K of the local moments within the dimerized 
Ti$^{3+}$-Ti$^{3+}$ pairs, whereas an inter-pair coupling is only 
of $\approx 40 $\,K.
This is in a good agreement with small and temperature independent 
magnetic susceptibility in the low-temperature phase of Ti$_4$O$_7$.

%%%%%%%%%%%%%%%%%%%%%%%%%%%%%%%%%%%%%%%%%%%%%%%%%%%%%%%%%%%%%%%%%%%%%%%%%%%%%%%%%
% Computational details
%%%%%%%%%%%%%%%%%%%%%%%%%%%%%%%%%%%%%%%%%%%%%%%%%%%%%%%%%%%%%%%%%%%%%%%%%%%%%%%%%

\section{Computational details}
\label{sec:details}

The present band-structure calculations have been performed for the 
low-temperature triclinic structure of Ti$_4$O$_7$. \cite{LM84} 
The $P\bar{1}$ unit cell used in the calculations
was constructed from the translation vectors of the original $I\bar{1}$ 
cell with $a=5.626$ \AA, $b=7.202$ \AA, $c=20.2608$ \AA, 
$\alpha=67.90^\circ$, $\beta=57.69^\circ$, and 
$\gamma=109.68^\circ$ found at 115\,K. 
%Its unit cell contains two formula units. 
The radii of muffin-tin spheres were taken as 
$R_{\rm Ti1-4}=2.27$ a.u., $R_{\rm O1,O3,O4-6}=1.78$ a.u., and 
$R_{\rm O2,O7}=1.66$ a.u. Fifteen kinds of empty spheres were 
introduced to fill up the inter-atomic space. For simplicity 
we neglect small spin-orbit coupling and consider only a 
collinear spin case.

%%%%%%%%%%%%%%%%%%%%%%%%%%%%%%%%%%%%%%%%%%%%%%%%%%%%%%%%%%%%%%%%%%%%%%%%%%%%%%%%%
% LSDA band structure
%%%%%%%%%%%%%%%%%%%%%%%%%%%%%%%%%%%%%%%%%%%%%%%%%%%%%%%%%%%%%%%%%%%%%%%%%%%%%%%%%

\section{LSDA band structure}
\label{sec:lsda}

Our LSDA band structure calculations for the low-temperature $P\bar{1}$ 
structure confirmed the results of the previous work.\cite{ESE04} 
The LSDA gives a nonmagnetic metallic solution with substantial charge 
separation between crystallographically independent Ti(1)/Ti(3) and 
Ti(2)/Ti(4) cations. The lower part of the valence band below -3\,eV is 
predominantly formed by O $2p$ states with a bonding hybridization 
with Ti $3d$ states. Crystal field splitting of the latter is roughly 
of 2.5\,eV. Ti $t_{2g}$ states form the group of bands at and up to 
2 eV above the Fermi energy whereas Ti $e_g$ states give a predominant 
contribution to the bands between 2.5 and 4.5\,eV. Within the $ t_{2g} $ 
group of bands the symmetry inequivalence of Ti(1)/Ti(3) and Ti(2)/Ti(4) 
sites leads to substantial $t_{2g}$ charge separation between these two 
groups of Ti atoms. In addition, an analysis of the partial density of 
states reveals significant bonding-antibonding splitting of $d_{xy}$ (in 
local cubic frame) states of about 1.5\,eV for Ti(1)/Ti(3) cations, 
whereas Ti(2)/Ti(4) cations show a relatively weak substructure. This 
substantial bonding-antibonding splitting of Ti(1)/Ti(3) $t_{2g}$ 
states agrees well with the concept of formation of Ti$^{3+}$-Ti$^{3+}$ 
spin-singlet pairs proposed earlier by Marezio.\cite{MMD73,LSC76}
However, the LSDA calculations fail to reproduce an insulating 
spin-singlet ground state of the low-temperature phase of Ti$_4$O$_7$. 
Apparently, the electron-electron correlations, mainly in the $3d$ shell 
of Ti cations, play a significant role.

%%%%%%%%%%%%%%%%%%%%%%%%%%%%%%%%%%%%%%%%%%%%%%%%%%%%%%%%%%%%%%%%%%%%%%%%%%%%%%%%%
% LDA+$U$ results and charge ordering
%%%%%%%%%%%%%%%%%%%%%%%%%%%%%%%%%%%%%%%%%%%%%%%%%%%%%%%%%%%%%%%%%%%%%%%%%%%%%%%%%

\section{LDA+$U$ results and charge ordering}
\label{sec:co}

\begin{figure}[tbp!]
\centerline{\includegraphics[width=0.45\textwidth,clip]{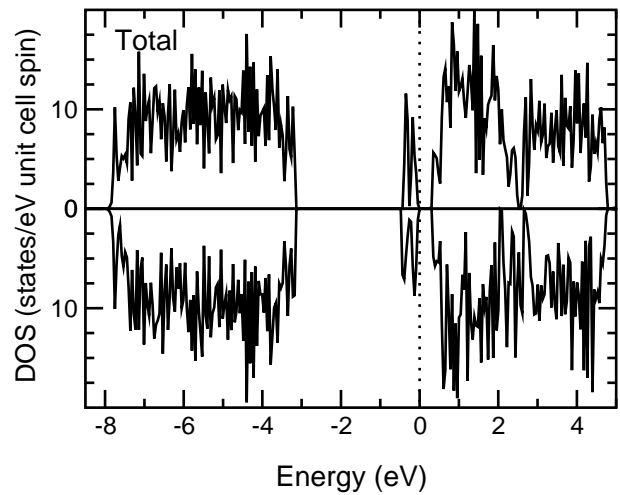}}
\caption{\label{fig:ldau}
The total DOS obtained from LDA+$U$ calculations with $U$=3.0\,eV and $J$=0.8\,eV  
for the low-temperature $P\bar{1}$ phase of Ti$_4$O$_7$. The top of the valence 
band is shown by dotted lines.}
\end{figure}

In order to take into account strong electronic correlations in Ti $3d$ 
shell we perform LDA+$U$ calculations for Ti$_4$O$_7$ in the low-temperature 
$P\bar{1}$ structure. 
In our calculations we use Coulomb intereaction parameter $U= 3.0$\,eV and
exchange coupling $J=0.8$\,eV taken in agreement with previous constrained 
LDA calculations.\cite{Ucalc}
The LDA+$U$ calculations result in a charge and orbitally 
ordered insulator with an energy gap of 0.29\,eV (see Fig.~\ref{fig:ldau}). 
This is in a strong contrast to the metallic solution with a substantial charge 
disproportionation between crystallographically inequivalent 
Ti(1)/Ti(3) and Ti(2)/Ti(4) cations obtained by LSDA and in 
a reasonably good agreement with an experimental gap value of 0.25\,eV.
Note, however, that the charge and orbital order pattern remains exactly 
the same for $U$ in the range 2.5-4.5\,eV, whereas the energy gap increases 
considerably up to 1.12\,eV for $U$=4.5\,eV. This remarkable increase of the 
gap value is accompanied by the enhancement of the spin magnetic moment from 
0.56 up to 0.8 $\mu_B$ per 3+ Ti(1)/Ti(3) cation as $U$ is increased from 2.5 
to 4.5\,eV. 

In addition, we perform LDA+$U$ calculations for high-temperature metallic 
phase of Ti$_4$O$_7$. In particular for $U$ of 2.5\,eV a metallic self-consistent 
solution with substantial density of states (76 states/Ry) at the Fermi 
level has been found, whereas for $U$ of 3\,eV the LDA+$U$ solution becomes 
unstable but remains metallic. With further increase of the $U$ value 
the metallic solution collapses into insulating one.

%%%%%%%%%%%%%%%%%%%%%%%%%%%%%%%%%%%%%%%%%%%%%%%%%%%%%%%%%%%%%%%%%%%%%%%%%%%%%%%%%
%%%%%%%%%%%%%%%%%%%%%%%%%%%%%%%%%%%%%%%%%%%%%%%%%%%%%%%%%%%%%%%%%%%%%%%%%%%%%%%%%

\begin{figure}[tbp!]
\centerline{\includegraphics[width=0.45\textwidth,clip]{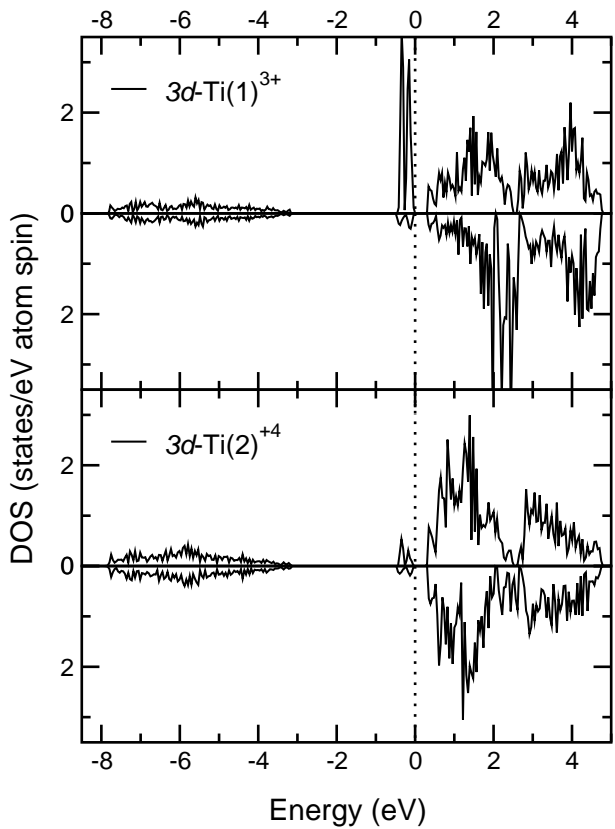}}
\caption{\label{fig:ldau-pdos}
The partial DOS for Ti(1)$^{3+}$ and Ti(2)$^{4+}$ cations are shown. 
The gap value of 0.29\,eV was obtained by LDA+$U$ with $U$=3.0\,eV 
and $J$=0.8\,eV. The Fermi level is shown by dotted line.}
\end{figure}

After self-consistency was achieved four crystallographically independent Ti 
atoms are split out in two subgroups in respect to the spin magnetic moment 
per Ti site: Ti(1)/Ti(3) with a moment of 0.66/-0.67 $\mu_B$, respectively, 
and Ti(2)/Ti(4) with 0.04/-0.02 $\mu_B$.  Thus, one of $t_{2g}$ 
majority/minority spin states of Ti(1)/Ti(3) becomes occupied $(d^1)$, 
whereas all other $t_{2g}$ states are pushed by strong Coulomb interaction 
above the Fermi level. In contrast, all $t_{2g}$ states of Ti(2) and 
Ti(4) are almost depopulated $(d^0)$ and form bands up to 2.5\,eV above 
the Fermi level. The occupied Ti(1)/Ti(3) states are strongly localized 
and form a prominent structure with a band width of 0.25\,eV 
just below the Fermi level (see Fig.~\ref{fig:ldau-pdos}).  The strong 
Coulomb interaction does not affect much the empty Ti $e_g$ states, 
which give predominant contribution between 2.5 and 4.5\,eV.
The obtained magnetic structure is almost antiferromagnetic with 
the spin magnetic moments within Ti(1)$^{3+}$-Ti(3)$^{3+}$ as well as
Ti(2)$^{4+}$-Ti(4)$^{4+}$ pairs being of the same magnitude with opposite
sign.  

%%%%%%%%%%%%%%%%%%%%%%%%%%%%%%%%%%%%%%%%%%%%%%%%%%%%%%%%%%%%%%%%%%%%%%%%%%%%%%%%%
%%%%%%%%%%%%%%%%%%%%%%%%%%%%%%%%%%%%%%%%%%%%%%%%%%%%%%%%%%%%%%%%%%%%%%%%%%%%%%%%%

%%%%%%%%%%%%%%%%%%%%%%%%%%%%%%%%%%%%%%%%%%%%%%%%%%%%%%%%%%%%%%%%%%%%%%%%%%%%%%%%%
% LSDA+U, order parameters
%%%%%%%%%%%%%%%%%%%%%%%%%%%%%%%%%%%%%%%%%%%%%%%%%%%%%%%%%%%%%%%%%%%%%%%%%%%%%%%%%

An analysis of occupation matrices of Ti(1)$^{3+}$/Ti(3)$^{3+}$ majority/minority
$3d$ spin states confirms substantial charge disproportionation within the Ti $3d$ 
shell. As shown in Table~\ref{tab:occ}, one of the $t_{2g}$ states of Ti$^{3+}$ 
cations $(d^1)$ is occupied with the occupation number of 0.74, whereas the 
remaining two $t_{2g}$ orbitals have a significantly smaller population of about 
0.08. Thus, according to Ref.~\onlinecite{LYA04} we define an orbital order 
parameter as the largest difference between 3+ and 4+ Ti $t_{2g}$ populations 
which amounts to 66\% of ideal ionic charge ordering model. The orbital order 
parameter clearly shows the existence of substantial charge disproportionation 
in the Ti 3$d$ shell of Ti$_4$O$_7$ which is remarkable because of the complete
lack of the total charge separation (see column $q$ in Table~\ref{tab:occ})
between 3+ and 4+ Ti cations. The occupation matrices analysis shows that 
the change of the $t_{2g}$ occupations is very efficiently screened by the 
rearrangement of the other Ti electrons. A significant portion of the screening 
charge is provided by Ti $e_g$ states due to formation of relatively strong 
$\sigma$ bonds with O $2p$ states, which results in appreciable contribution of 
the former to the occupied part of the valence band. Ti $4s$ and $4p$ states 
give additional contributions to the screening of the difference in $t_{2g}$
occupations which leads to complete loss of the disproportionation between 
the charges at 3+ and 4+ Ti sites.

%%%%%%%%%%%%%%%%%%%%%%%%%%%%%%%%%%%%%%%%%%%%%%%%%%%%%%%%%%%%%%%%%%%%%%%%%%%%%%%%%
\begin{table}[tbp!]
\caption{\label{tab:occ}Total $(q)$ and $l$-projected $(q_{s,p,d})$ charges, 
magnetic moments $(M)$, 
and occupation of the most populated $t_{2g}$ orbitals $(n)$ calculated for
inequivalent Ti atoms in the low-temperature $P\bar{1}$ phase of Ti$_4$O$_7$.}
\begin{ruledtabular}
\begin{tabular}{lccccccc}
Ti ion & $q$ &$q_s$ &$q_p$ &$q_d$ & $M$ ($\mu_{\text{B}}$) & $t_{2g}$ orbital & $n$ \\
\hline
Ti(1)$^{3+}$  & 2.27 & 0.18 & 0.27 & 1.83 & 0.66 & $d_{xy\uparrow}$ & 0.74 \\
Ti(2)$^{4+}$  & 2.22 & 0.22 & 0.33 & 1.68 & 0.04 &  & 0.08 \\
Ti(3)$^{3+}$  & 2.16 & 0.18 & 0.25 & 1.74 &  -0.67 & $d_{xy\downarrow}$ & 0.73 \\
Ti(4)$^{4+}$  & 2.16 & 0.21 & 0.33 & 1.62 &  -0.02 &  & 0.07 \\
\end{tabular}
\end{ruledtabular}
\end{table}
%%%%%%%%%%%%%%%%%%%%%%%%%%%%%%%%%%%%%%%%%%%%%%%%%%%%%%%%%%%%%%%%%%%%%%%%%%%%%%%%%

%%%%%%%%%%%%%%%%%%%%%%%%%%%%%%%%%%%%%%%%%%%%%%%%%%%%%%%%%%%%%%%%%%%%%%%%%%%%%%%%%
% LSDA+U, orbitals
%%%%%%%%%%%%%%%%%%%%%%%%%%%%%%%%%%%%%%%%%%%%%%%%%%%%%%%%%%%%%%%%%%%%%%%%%%%%%%%%%

\begin{figure}[tbp!]
\centerline{\includegraphics[width=.45\textwidth,clip]{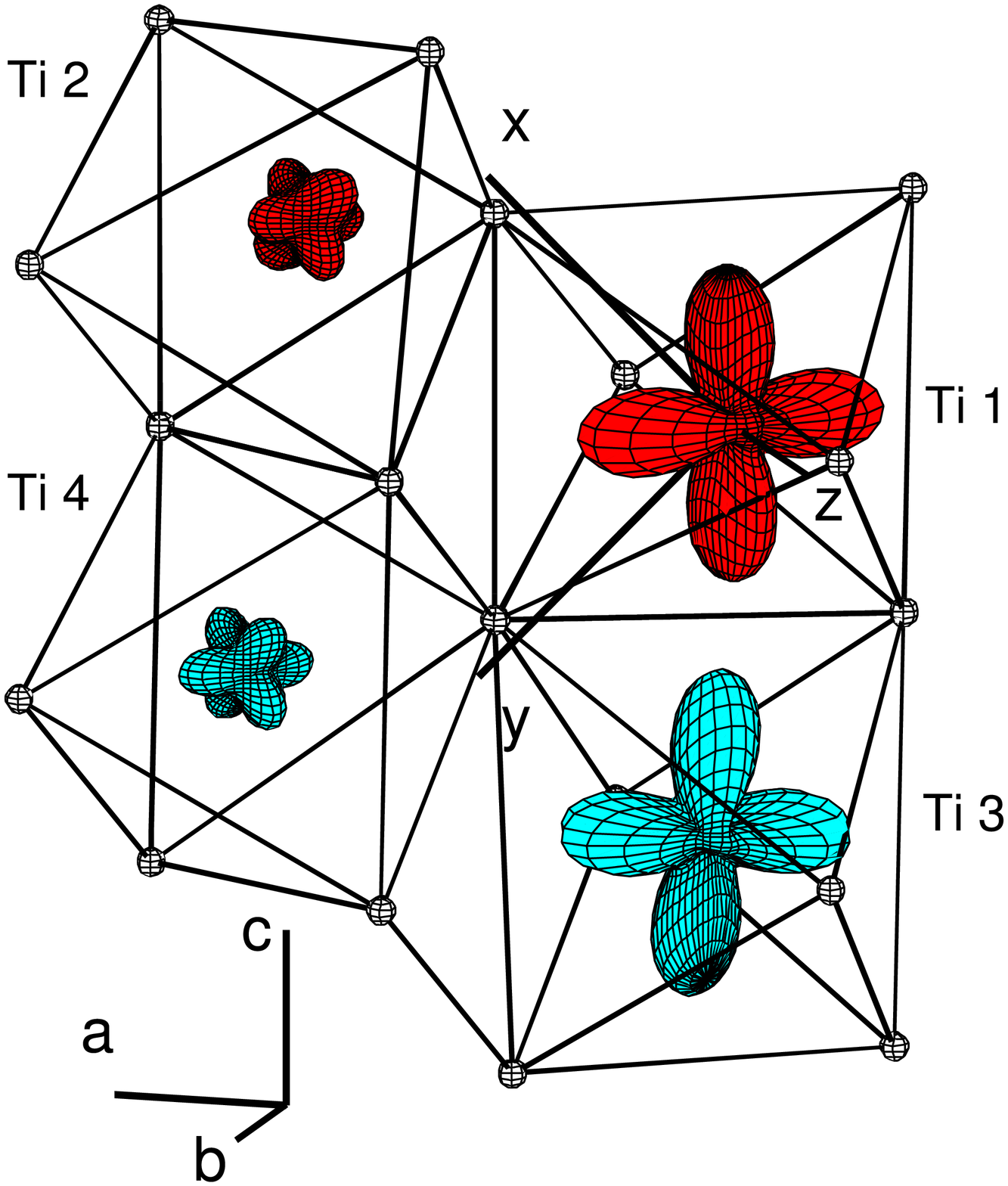}}
\caption{\label{fig:orb} 
(color online)
Structure of Ti$_4$O$_7$ showing the angular distribution of the majority 
and minority spin $3d$ electron density of Ti cations. Red and cyan (light and 
dark, respectively, on the black and white image) orbitals correspond to the 
majority and minority $3d$ spin states, respectively. Oxygen atoms are shown 
by small spheres. The size of orbital corresponds to its occupancy. X-Y-Z 
coordinate system corresponds to the local cubic frame.}
\end{figure}

The occupied $t_{2g}$ Ti$^{3+}$ states are predominantly of $d_{xy}$ 
character in the local cubic frame (according to that we later mark the orbital 
as $d_{xy}$ orbital). This is illustrated in Fig.~\ref{fig:orb}, which shows the 
angular distribution of the majority and minority spin $3d$ electron density 
of Ti cations, marked by red and cyan color (or light and dark on the black 
and white image), respectively.\cite{chargedens} 
Since Ti(1)$^{3+}$ and Ti(3)$^{3+}$ cations are antiferromagnetically coupled,
the obtained ferro-orbital order is consistent with the formation of a bonding 
spin-singlet 
state from the $d_{xy}$ orbitals of two neighboring Ti(1) and Ti(3) sites.
The orientation of occupied Ti$^{3+}$ t$_{2g}$ orbitals is consistent with
the largest average Ti-O distance in the plane of t$_{2g}$ orbitals. 
As shown in Table~\ref{tab:dist} the average Ti(1)-O distance (2.061 \AA) 
in the plane of $d_{xy}$ orbital is considerably larger than average distances 
in the other two $yz$ and $zx$ planes (2.032 and 2.045 \AA, respectively).
The same is also true for the Ti(3) cation but in this case the variation 
of the average Ti(3)-O distances is much smaller (2.047 vs 2.041 and 2.042 \AA) 
and, as a consequence, the out-of-plane rotation of the occupied $t_{2g}$ 
minority spin orbital is stronger.

%%%%%%%%%%%%%%%%%%%%%%%%%%%%%%%%%%%%%%%%%%%%%%%%%%%%%%%%%%%%%%%%%%%%%%%%%%%%%%%%%
%%%%%%%%%%%%%%%%%%%%%%%%%%%%%%%%%%%%%%%%%%%%%%%%%%%%%%%%%%%%%%%%%%%%%%%%%%%%%%%%%

%%%%%%%%%%%%%%%%%%%%%%%%%%%%%%%%%%%%%%%%%%%%%%%%%%%%%%%%%%%%%%%%%%%%%%%%%%%%%%%%%
\begin{table}[tbp!]
\caption{\label{tab:dist}The averaged Ti-O distances in the plane of 
$t_{2g}$ orbitals $(d_{\rm orb.})$ and in the oxygen octahedra $(d_{\rm av.})$ 
for $P\bar{1}$ structure of Ti$_4$O$_7$. The occupied orbital of the $3d^1$ 
Ti(1) and Ti(3) 3+ cations is predominantly of $d_{xy}$ character.}
\begin{ruledtabular}
\begin{tabular}{lccc}
Ti atom & orbital & $d_{\mathsf{orb.}}$ (\AA) & $d_{\mathsf{av.}}$ (\AA)\\
\hline
Ti(1)                 & $d_{xy}$   & 2.061 & 2.046  \\
                      & $d_{yz}$   & 2.032 &        \\
                      & $d_{zx}$   & 2.045 &        \\              
Ti(2)                 & $d_{xy}$   & 2.012 & 2.000  \\
                      & $d_{yz}$   & 1.976 &        \\
                      & $d_{zx}$   & 2.013 &        \\
Ti(3)                 & $d_{xy}$   & 2.047 & 2.043  \\
                      & $d_{yz}$   & 2.041 &        \\
                      & $d_{zx}$   & 2.042 &        \\
Ti(4)                 & $d_{xy}$   & 1.973 & 1.977  \\
                      & $d_{yz}$   & 1.976 &        \\
                      & $d_{zx}$   & 1.981 &        \\
\end{tabular}
\end{ruledtabular}
\end{table}
%%%%%%%%%%%%%%%%%%%%%%%%%%%%%%%%%%%%%%%%%%%%%%%%%%%%%%%%%%%%%%%%%%%%%%%%%%%%%%%%%

In addition, hopping matrix elements were evaluated via Fourier 
transformation from reciprocal to real space of the Ti $t_{2g}$ 
LDA Wannier Hamiltonian.\cite{Wannier} Remarkably, for the 
low-temperature phase the Ti(1)-Ti(3) 
intra-pair $d_{xy}$-$d_{xy}$ hopping matrix element
is found to be of 0.61 eV, whereas all other hoppings are
3-4 times smaller. This strong inhomogeneity of the hopping matrix elements
disappears in the 
high-temperature phase. Thus, according to our calculations 
hopping elements in the high-temperature phase are 0.23, 
0.21, 0.39, and 0.33 eV between 1-3, 2-4, 3-3, and 4-4 Ti sites, 
respectively. 

Estimation of exchange interaction parameters via the variation
of the ground state energy with respect to the magnetic moment 
rotation angle\cite{LAZ95,AAL97} results in a strong 
antiferromagnetic coupling of -1696\,K between Ti(1)$^{3+}$ 
and Ti(3)$^{3+}$ cations.\cite{exchanges} All other couplings 
are two orders of magnitudes smaller. This indicates 
a possible formation 
of the spin-singlet pairs via direct antiferromagnetic exchange 
between neighboring Ti(1) and Ti(3) sites. The contribution of 
the superexchange via O $p$ orbitals to the Ti(1)-Ti(3) exchange 
coupling is found to be negligible. This was verified by calculating 
the exchange coupling constants with the sub-blocks 
of the LMTO Hamiltonian responsible for the Ti-O hybridization being 
set to zero. This calculation gave qualitatively same results for 
the exchange constants although the possibility for the
superexchange via O $p$ orbitals was eliminated.

%%%%%%%%%%%%%%%%%%%%%%%%%%%%%%%%%%%%%%%%%%%%%%%%%%%%%%%%%%%%%%%%%%%%%%%%%%%%%%%%%
% Summary and conclusions
%%%%%%%%%%%%%%%%%%%%%%%%%%%%%%%%%%%%%%%%%%%%%%%%%%%%%%%%%%%%%%%%%%%%%%%%%%%%%%%%%

\section{Summary and conclusions}
\label{sec:sum}

In summary, in the present LDA+$U$ study of the low-temperature $P\bar{1}$ phase 
of Ti$_4$O$_7$ we found a charge ordered insulating solution with an energy gap of 0.29\,eV. 
The total $3d$ charge separation is small (less than 0.14), whereas
the orbital order parameter defined as the difference between $t_{2g}$ occupancies 
of Ti$^{3+}$ and Ti$^{4+}$ cations is large and gives direct evidence for charge 
ordering. Ti $4s$ and $4p$ states give a strong contribution to 
the static ``screening'' of the total $3d$ charge separation. This effective 
charge screening leads to complete loss of the disproportionation between the 
charges at 3+ and 4+ Ti sites.
The occupied $t_{2g}$ states of Ti$^{3+}$ cations
are predominantly of $d_{xy}$ character (in the local cubic frame) and form a 
spin-singlet molecular orbital via strong direct antiferromagnetic exchange 
coupling between neighboring Ti(1) and Ti(3) sites of $ J \approx 1700 $\,K, 
whereas the role of superexchange is found to be negligible. 
This is in a good agreement with small and temperature independent 
magnetic susceptibility in the low-temperature phase of Ti$_4$O$_7$.

%%%%%%%%%%%%%%%%%%%%%%%%%%%%%%%%%%%%%%%%%%%%%%%%%%%%%%%%%%%%%%%%%%%%%%%%%%%%%%%%%
% Acknowledgements
%%%%%%%%%%%%%%%%%%%%%%%%%%%%%%%%%%%%%%%%%%%%%%%%%%%%%%%%%%%%%%%%%%%%%%%%%%%%%%%%%

\section{Acknowledgements}
\label{sec:ack}

We are grateful to K.~Schwarz, P.~Blaha, P.~Fulde, and D.~Vollhardt for helpful 
discussions. The present work was supported in part by RFFI Grant No. 04-02-16096, 
No. 03-02-39024, NWO 047.016.005, and by the Sonderforschungsbereich 484 
of the Deutsche Forschungsgemeinschaft (DFG).


\begin{thebibliography}{34}

\bibitem{Coey04} M. Coey, Nature {\bf 430}, 155 (2004).

\bibitem{Rev01} Several reviews of research on the Verwey transition up to
  1980 are contained in the special issue of Philos. Mag. B {\bf 42} (1980).
	
\bibitem{V39} E. J. W. Verwey, Nature (London) {\bf 144}, 327 (1939).

\bibitem{VHR47} E. J. W. Verwey, P. W. Haayman, and F. C. Romeijan, 
J. Chem. Phys. {\bf 15}, 181 (1947).

\bibitem{LYA04} I. Leonov, A. N. Yaresko, V. N. Antonov, M. A. Korotin, and V. I. Anisimov,
Phys. Rev. Lett. {\bf 93}, 146404 (2004).

\bibitem{JGH04} Horng-Tay Jeng, G. Y. Guo, and D. J. Huang, 
Phys. Rev. Lett. {\bf 93}, 156403 (2004).

\bibitem{screening} 
Here and in the following we assume a redistribution of charge between Ti 
$t_{2g}$ and other states using term screening.

\bibitem{LYA05} I. Leonov, A. N. Yaresko, V. N. Antonov, J. P. Attfield, and V. I. Anisimov,
Phys. Rev. B {\bf 72}, 014407 (2005).

\bibitem{ACP92} J. P. Attfield, J. F. Clarke, and D. A. Perkins, 
Physica B {\bf 180-181}, 581 (1992).

\bibitem{ABRM98} J. P. Attfield, A. M. T. Bell, L. M. Rodriguez-Martinez, J. M. Greneche, 
R. J. Cernik, J. F. Clarke, and D. A. Perkins, Nature {\bf 396}, 655 (1998).

\bibitem{ABRM99} J. P. Attfield, A. M. T. Bell, L. M. Rodriguez-Martinez, J. M. Greneche, 
R. Retoux, M. Leblanc, R. J. Cernik, J. F. Clarke, and D. A. Perkins, 
J. Mater. Chem. {\bf 9}, 205 (1999).

\bibitem{V7O13} To our knowledge of all the V$_n$O$_{2n-1}$ compounds only
V$_7$O$_{13}$ does not exhibit a metal-insulator transition.

\bibitem{schwingen03}
U.\ Schwingenschl\"ogl, V.\ Eyert, and U.\ Eckern, Europhys.\ Lett.\ {\bf 61}, 361 (2003),
{\it ibid.} {\bf 64}, 682 (2003), and 
U.\ Schwingenschl\"ogl and V.\ Eyert, Ann.\ Phys.\ (Leipzig) {\bf 13}, 475 (2004).

\bibitem{wentz94a}
R.\ M.\ Wentzcovitch, W.\ W.\ Schulz, and P.\ B.\ Allen,
Phys.\ Rev.\ Lett.\ {\bf 72}, 3389 (1994).

\bibitem{KSA02}
M. A. Korotin, N. A. Skorikov, V. I. Anisimov,
The Physics of Metals and Metallography {\bf 94}, 17, 2002.

\bibitem{eyert02b}
V.\ Eyert, Ann.\ Phys.\ (Leipzig) {\bf 11}, 650 (2002).

\bibitem{eyert00}
V.\ Eyert, R.\ Horny, K.-H.\ H\"ock, and S.\ Horn,
J.\ Phys.: Cond.\ Matt.\  {\bf 12}, 4923 (2000).

\bibitem{eyert02a}
V.\ Eyert, Europhys.\ Lett.\ {\bf 58}, 851 (2002).

\bibitem{biermann05} 
S.\ Biermann, A.\ Poteryaev, A.\ I.\ Lichtenstein, A.\ Georges, 
Phys.\ Rev.\ Lett.\ {\bf 94}, 026404 (2005).

\bibitem{BF69} R. F. Bartholomew and D. R. Frankl, Phys. Rev. {\bf 187}, 828 (1969).

\bibitem{LSC76} S. Lakkis, C. Schlenker, B. K. Chakraverty, R. Buder, and M. Marezio,
Phys. Rev. B {\bf 14}, 1429 (1976).

\bibitem{MMD72} M. Marezio, D. B. McWhan, P. D. Dernier, and J. P. Remeika
Phys. Rev. Lett. {\bf 28}, 1390 (1972).

\bibitem{MD71} M. Marezio and P. D. Dernier
J. Soli State Chem. {\bf 3}, 340 (1971).

\bibitem{HM79} J. L. Hodeau and M. Marezio
J. Soli State Chem. {\bf 29}, 47 (1979).

\bibitem{LM84} Y. LePage and M. Marezio, J. Solid State Chem. 53, 13 (1984)

\bibitem{ESE04} V. Eyert, U. Schwingenschl\"ogl, and U. Eckern, 
Chem. Phys. Lett. {\bf 390}, 151 (2004).

\bibitem{ole} O. K. Andersen, Phys. Rev. B {\bf 12}, 3060 (1975).

\bibitem{AZA91} V. I. Anisimov, J. Zaanen, and O. K. Andersen, 
Phys. Rev. B {\bf 44}, 943 (1991).

\bibitem{LAZ95} A. I. Liechtenstein, V. I. Anisimov, and J. Zaanen, 
Phys. Rev. B {\bf 52}, R5467 (1995).

\bibitem{MMD73} M. Marezio, D. B. McWhan, P. D. Dernier, and J. P. Remeika
J. Solid State Chem. {\bf 6}, 213 (1973).

\bibitem{Ucalc} S. V. Streltsov, A. S. Mylnikova, A. O. Shorikov, Z. V. Pchelkina, 
D. I. Khomskii, and V. I. Anisimov, Phys. Rev. B {\bf 71}, 245114 (2005);
I. Solovyev, N. Hamada, and K. Terakura, Phys. Rev. B {\bf 53}, 7158 (1996), and
T. Mizokawa and A. Fujimori, Phys. Rev. B {\bf 54}, 5368 (1996).

\bibitem{chargedens} 
The distribution is calculated according to 
$\rho(\theta,\phi)=\sum_{m,m'}n_{m,m'}~Y^\ast_{m}(\theta,\phi)~Y_{m'}(\theta,\phi)$,
where $n_{m,m'}$ is the occupation matrix of $3d$ majority states of Ti(1) and
$3d$ minority states of Ti(3) cations. The occupation matrices were calculated by 
the LDA+$U$ with $U$= 3.0\,eV, $J$= 0.8\,eV for the low-temperature $P\bar{1}$ phase 
of Ti$_4$O$_7$. $Y_{m}(\theta,\phi)$ denotes corresponding spherical harmonics. 

\bibitem{Wannier}
V. I. Anisimov, D. E. Kondakov, A. V. Kozhevnikov, I. A. Nekrasov, 
Z. V. Pchelkina, J. W. Allen, S.-K. Mo, H.-D. Kim, P. Metcalf, 
S. Suga, A. Sekiyama, G. Keller, I. Leonov, X. Ren, 
and D. Vollhardt, Phys. Rev. B {\bf 71}, 125119 (2005).

\bibitem{AAL97} V. I. Anisimov, F. Aryasetiawan, and A. Lichtenstein, 
J. Phys.: Condens. Matter {\bf 9}, 767 (1997).

\bibitem{exchanges} 
The exchange coupling parameter $J$ represents the effective pair 
exchange interaction between Ti atoms with effective Heisenberg Hamiltonian 
$H = - \sum_{i > j} J_{ij} S_i \cdot S_j$, where $S_i$ and $S_j$ are spin 
magnetic moment vectors at site $i$ and $j$. Positive (negative) values of $J$ 
correspond to the ferromagnetic (antiferromagnetic) coupling between sites.

\end{thebibliography}
\end{document}